# Pressure-induced reversible phase transition and amorphization of $CH_3NH_3PbI_3$

*Kai Wang,[†] Ran Liu,[†] Yuancun Qiao,[†] Jinxing Cui,[†] Bo Song,[‡] Bingbing Liu,[†] and Bo Zou,\*[†]*

[†]State Key Laboratory of Superhard Materials, Jilin University, Changchun 130012, P. R. China

[‡]Suzhou Key Laboratory of Macromolecular Design and Precision Synthesis, Jiangsu Key Laboratory of Advanced Functional Polymer Design and Application, College of Chemistry, Chemical Engineering and Materials Science, Soochow University, Suzhou 215123, China.

\*Corresponding author. E-mail: zoubo@jlu.edu.cn, Tel: 86-431-85168882

**Abstract**

Recent advances of highly efficient solar cells based on organic-inorganic halide perovskites have triggered intense research efforts to establish the fundamental properties of these materials. In this work, we utilized diamond anvil cell to investigate the pressure-induced structural and electronic transformations in methylammonium lead iodide ($CH_3NH_3PbI_3$) up to 7 GPa at room temperature. The synchrotron X-ray diffraction experiment show that the sample transformed from tetragonal to orthorhombic phase at 0.3 GPa and amorphized above 4 GPa. Further high pressure IR spectroscopy experiments illustrated the high pressure behavior of organic $(CH_3NH_3)^+$ cations. We also analyzed the pressure dependence of the band gap energy based on the optical absorption and photoluminescence (PL) results. Moreover, all the observed changes were fully reversible when the pressure was completely released. Our *in situ* high pressure studies provide essential information for the intrinsic properties and stability of organic-inorganic halide perovskites, which significantly affect the performance of perovskite solar cells.

**Introduction**

Organic–inorganic halide perovskites have drawn considerable interest recently owing to their outstanding photovoltaic performance in solar cells. With a span of four years, the power conversion efficiencies of organic–inorganic halide perovskite solar cells have been significant improved from 3% to over 20%.[1,2] Among them, methylammonium lead iodide ($CH_3NH_3PbI_3$) is the most brilliant light harvester because of its appropriate band gap energy and high absorption coefficient.[3] However, the fundamental structural, optical and electronic properties of $CH_3NH_3PbI_3$ are not fully understood so far.[4] One essential issue is to investigate the chemical stability and structure-property relationships at various surrounding environment (e.g., moisture, light, temperature, and pressure), which has considerable appeal for practical application. Hydrostatic pressure is a thermodynamics variable which can effectively tune crystal structure and electronic properties. A systematic high pressure study of $CH_3NH_3PbI_3$ will provides further insight into the intrinsic characteristics and thus contributes to elucidating its exceptional photovoltaic properties.

$CH_3NH_3PbI_3$ crystals have perovskite-type structure ($AMX_3$), in which methylammonium ($CH_3NH_3^+$) locates in A site and Pb at the M site. The $[PbI_6]^{4-}$ octahedra are corner-connected to form a three-dimensional (3D) network and the $CH_3NH_3^+$ cations are filled in the vacancies of this framework (Figure 1). Interestingly, previous calculations indicate that the band gap energy of $CH_3NH_3PbI_3$ is mainly governed by the $[PbI_6]^{4-}$ network and the photoelectric properties are highly associated with its crystal structure.[5-7] At ambient conditions, the $CH_3NH_3PbI_3$ crystal belongs to tetragonal symmetry (space group, *I4cm*) with a band gap at about 1.6 eV and a light absorption spectrum in the whole visible region.[8] The tetragonal phase will transform to a pseudocubic phase (tetragonal, *P4mm*) with increasing pressure at 330 K or transform to an orthorhombic phase (space group, *Pnma*) on cooling below 161 K. These phase transitions can be mainly attributed to the tilting of $[PbI_6]^{4-}$ octahedra which largely affected band gap energies.[9] Early high pressure researchers have found a new high pressure phase (above 0.09 GPa) of $CH_3NH_3PbI_3$ using DAT and dielectric techniques, respectively.[10,11] However, there is no structural information and optical properties of this high pressure phase are available in literatures. Using diamond anvil cells, Matsuishi *et al.* studied high pressure effect on the electronic structure and the excitonic states of $CH_3NH_3PbBr_3$.[12] Then Swainson and Zhao *et al.* confirmed the phase transition from $Pm\bar{3}m$ to $Im\bar{3}$ and found a pressure-induced amorphization around 2 GPa.[13,14] The pressure-induced conductivity and piezochromic of a 2-D Cu-Cl hybrid perovskites have also been reported recently.[15] However, up to now, only limited works have been conducted about the high pressure behavior of organic-inorganic halide perovskites and research in this field is still infant.

Herein, we have studied the evolution of $CH_3NH_3PbI_3$ crystal structure and optical properties as a function of pressure using diamond anvil cells. The crystal structure of new high pressure phase and phase transition mechanism were suggested based on high intensity synchrotron XRD and Rietveld refinement. The variation of band gap energy was discussed based on high pressure optical absorption and photoluminescence (PL) spectra. The primary goal of this study is to provide a better understanding of the structural stability and the nature of high efficiency perovskite solar cells.

**Experimental Section**

The $CH_3NH_3PbI_3$ sample (purity > 99%) was purchased from Xi'an Polymer Light Technology Corp. and used without further purification. The XRD patterns confirmed that the obtained sample is pure tetragonal phase. High pressure experiments were performed with a symmetric diamond anvil cell. The culet diameter of the diamond anvils was 0.4 mm. A SUS301 stainless steel gasket with a 0.15 mm diameter sample hole and an initial thickness of 0.04 mm was used. A small ruby ball was placed in the sample chamber for in situ pressure measurements by the standard ruby fluorescence technique.[16] Owing to the low phase transition pressure, we have employed three kinds of pressure transmitting medium for different experiments (argon, silicone oil, and KBr). All experiments were performed at room temperature.

High-pressure synchrotron ADXRD experiments were conducted at the 4W2 high-pressure station of the Beijing Synchrotron Radiation Facility. XRD data were recorded using imaging plate detector and transformed into one-dimensional XRD patterns with FIT2D program.[17] The high-pressure XRD results were then analyzed using the powder reflex module in the Materials Studio Software. High pressure PL and absorption spectra were performed using a fiber spectrometer (QE65000, Ocean Optics) A 532 nm single-mode DPSS laser was used as the light source for PL excitation. Optical images of the compressed samples were obtained using a camera (Canon EOS 5D Mark II) equipped on a microscope (Eclipse Ti-U, Nikon). High pressure IR measurements were conducted using a Bruker VERTEX 80v spectrometer attached to an IR microscope.

**Results and Discussion**

Figure 2 shows representative synchrotron XRD patterns of $CH_3NH_3PbI_3$ obtained at different pressures in the process of compression and decompression. The maximum pressure achieved in our experiment was up to 7 GPa. The first XRD pattern at 1 atm indicate that the initial crystal structure is the pure tetragonal phase with space group *I*4*cm*. As pressure increased, all the diffraction peaks shifted to higher angles with respect to the pressure-induced decrease in distance between atoms. However, there are many new diffraction peaks appeared suddenly (denoted by arrows in Figure 2) and some original peaks disappeared (such as 211 peak) completely at 0.3 GPa. We note that the intensity of these new Bragg peaks increases gradually with increasing pressure and the new set of diffraction pattern is finally formed at 1 GPa. From the evolution of high pressure XRD patterns , it is evident that the tetragonal phase of $CH_3NH_3PbI_3$ undergo a pressure-induced phase transition which occurred at 0.3 GPa and completed at 1 GPa. But there are no mixed phase states can be observed during the entire phase transition process. With further compression, apart from pressure-induced shifts, it is noteworthy that all the Bragg peaks gradually become broad. But in general, the diffraction pattern of new high pressure phase is maintained up to 4 GPa with no significant change. When the applied pressure exceeds 4 GPa, a broad background (at about 12°) originating from the diffuse scattering appeared and the intensity of all original peaks suddenly decreased and most of them disappeared with increasing pressure, indicating the occurrence of pressure-induced amorphization. Although some relative strong diffraction peaks coming from residual crystalline state can persisted up to 7 GPa, the highest pressure achieved in this study. On decompression, the amorphous state sample first converts to crystalline high pressure phase at 0.5 GPa and then returned to the initial tetragonal phase after complete release of pressure. The high pressure XRD results provide strong evidence for a reversible pressure-induced phase transition at 0.3 GPa and a reversible pressure-induced amorphization above 4 GPa of tetragonal $CH_3NH_3PbI_3$.

To better understand the pressure-induced phase transition, it is essential to know the crystal structure of high pressure phase. Except for the studied tetragonal phase, $CH_3NH_3PbI_3$ also have a high temperature pseudocubic phase (tetragonal, *P4mm*) and a low temperature orthorhombic phase (*Pnma*). However, the XRD pattern of high pressure phase is obvious difference from them. So we preformed Rietveld refinement of the XRD pattern at 1 GPa. The refined result is a orthorhombic structure (pseudo-cubic, a = 12.253(8) Å, b = 12.256(4) Å,c = 12.269(5) Å) with space group *Imm*2. As shown in Figure 3 (b), the simulated XRD pattern agrees well with the experimental XRD data and the refinement quality factors (Rwp = 4.28%, Rp = 3.14%) are satisfactory. It is important to note that the number and position of simulated Bragg reflections in excellent agreement with experiment peaks. Figure 3 (a) present the refined orthorhombic crystal structure. Compared to the original tetragonal structure, the unit cell of orthorhombic structure enlarged about two times (a ≈ $\sqrt{2}$ $a_0$, b ≈ $\sqrt{2}$ $b_0$, c ≈ $c_0$) and the $[PbI_6]^{4-}$ network distorted more severely. For example, the corner-connected $[PbI_6]^{4-}$ octahedra in tetragonal phase are arranged in linear chains along the c-axis (Figure 1 (b)) while in orthorhombic phase octahedra are tilted substantially away from linear arrangement. Moreover, the $[PbI_6]^{4-}$ octahedra are also distorted along the *b*-axis after phase transition (Figure 3 (a-iii)). So it is evident that the observed pressure-induced phase transition is mainly caused by the tilting and distortion of $[PbI_6]^{4-}$ octahedra. It is also interesting to compare the $[PbI_6]^{4-}$ octahedra network of high pressure orthorhombic structure with that of low temperature orthorhombic structure. They have similar topological features viewed along the *b* axis (Figure 3 (a-ii)) but have different arrangement viewed from the other two axes. However, further high pressure single-crystal XRD or neutron diffraction studies are essential to confirm our results and provide more accurate atomic positions.

The variation of the unit cell volumes as a function of pressure is displayed in Figure 4. As expected, the volume of tetragonal phase decreases gradually with increasing pressure. However, the unit cell volume undergoes a sudden reduction at 0.3 GPa, which corresponding to the observed phase transition. The apparent discontinuities in the pressure-volume curve suggesting that this pressure-induced phase transition is a first-order type. After phase transition, the unit cell volume of orthorhombic structure continuing to shrink with a faster rate than tetragonal phase. This pressure-volume data were fitted using the third-order Birch-Murnaghan equation of state yielded $B_0 = 15.3(6)$ GPa and $B_0' = 6.8(2)$ GPa. The low value of bulk modulus $B_0$ and its pressure derivative $B_0'$ demonstrates the high compressible nature of $CH_3NH_3PbI_3$ and related organic-inorganic halide perovskites.[18]

In order to investigate the high pressure behavior of organic cation ($CH_3NH_3^+$), which has small x-ray scattering cross section compared to Pb and I atoms, we performed high pressure IR experiment. The evolution of IR spectra as a function of pressure is present in Figure 5. The IR spectra of $CH_3NH_3PbI_3$ have two main bands in the 1350-1650 $cm^{-1}$ range and two distinguishable bands between 3000 and 3400 $cm^{-1}$, which are attributed to C-H and N-H bending modes and N-H stretching modes, respectively.[19,20] With increasing pressure, the bending modes are continuously redshift while the stretching modes exhibit at first a blueshift and then a redshift above 2 GPa. Considering the existence of hydrogen bonging between organic cations and inorganic framework, the redshift of the observed peaks can be attribute to the strengthen of N–H⋯I and C–H⋯I hydrogen bond under high pressure.[21,22] The transition from blueshift to redshift of N-H stretching modes is also indicating that the attractive interactions between hydrogen atoms and iodine atoms are gradually strengthening. We note that there is no obvious change of the IR spectra at 0.3 GPa, the phase transition pressure. This may be due to $CH_3NH_3^+$ cations located in the relative large Pb-I cage and weak interactions between them in low pressure range. Another notable feature is that most of the IR peaks tend to broaden with increasing pressure. The broadening of vibration peaks, especially after amorphization, can be mainly attributed to the distortion of relatively soft organic $CH_3NH_3^+$ cations under high pressure. The flexible organic cations act as templates for the P-I frameworks which leads to the structural memory effects as observed in the reversible amorphization.

Considering its outstanding photovoltaic performance, it is necessary to investigate the high pressure optical property of $CH_3NH_3PbI_3$. Figure 6 shows the high pressure optical micrographs in a diamond anvil cell. On compression, the opaque black sample gradually transformed to transparent red color above 4 GPa, indicating that the observed pressure-induced amorphization can significantly change its optical properties. Further high pressure optical absorption and PL spectra confirmed the results of optical micrographs, as shown in Figure 7. The absorption edges and PL peaks are in good agreement at all pressures. As pressure was increased, the absorption edge and PL peak move towards longer wavelength region of 9 nm up to 0.25 GPa. However, there is an abrupt blueshift of absorption edge and PL peak occurred at 0.3 GPa, following a further gradual blueshift up to 1 GPa, which are corresponding to the observed phase transition. With further increase of pressure, the absorption edge and PL peak tend to slight redshift again. According to reports in the literature,[9] the band gap energy is mainly determined by the $[PbI_6]^{4-}$ network and the organic cations have little influence on it. So the decrease of band gap can be attributed to the shrinkage of $[PbI_6]^{4-}$ octahedra while the increase of band gap in the 0.25-1 GPa region is governed by the symmetry decreases and the tilting of octahedral network increases in the process of the tetragonal-orthorhombic phase transition. Above 3 GPa, a two-step variation of the absorption edge and two PL peaks are observed. The original absorption edge and PL peak disappeared at about 4 GPa while the new absorption edge and PL peak at higher energies disappeared completely at 5 GPa. Compared to the high pressure XRD results, these significant variation changes can be attribute to the structural adjustment in the process of amorphization. Moreover, the disappearance of absorption edge and appearance of a broad absorption tail can reasonably explain the pressure-induced black to red transition of $CH_3NH_3PbI_3$ sample. Notably, the observed changes in optical micrographs, optical absorption and PL spectra completely revert to its original state because of the reversible pressure-induced structural changes.

**Conclusion**

In summary, we studied the structural and optical properties of organic-inorganic halide perovskite $CH_3NH_3PbI_3$ as a function of pressure. We observed a reversible pressure-induced phase transition at 0.3 GPa and identified the high pressure phase as orthorhombic structure (space group *Imm*2) with the help of Rietveld refinement. This phase transition can slightly increase the band gap energy due to symmetry lowering and octahedra tilting. On further compression, $CH_3NH_3PbI_3$ exhibits a reversible pressure-induced amorphization above 4 GPa, which significantly affect its photovoltaic properties. The disappearance of absorption edge of amorphous $CH_3NH_3PbI_3$ will reduces the absorption ability of solar radiation. These results provide useful insights into the structural stability and intrinsic properties of $CH_3NH_3PbI_3$, promoting its future applications in solar cell devices.

**Acknowledgments**

This work is supported by NSFC (Nos. 91227202, and 11204101), RFDP (No. 20120061130006), Changbai Mountain Scholars Program (No. 2013007), National Basic Research Program of China (No. 2011CB808200). Angle-dispersive XRD measurement was performed at beamline 4W2, Beijing Synchrotron Radiation Facility (BSRF) which is supported by Chinese Academy of Sciences (No. KJCX2-SW-N20, KJCX2-SW-N03).

**Figure Captions**

**Figure 1.** The crystal structure and of the tetragonal $CH_3NH_3PbI_3$ viewed along different directions. I, Pb, N, C, and H atoms are depicted in purple, gray, blue, brown, and pink, respectively.

**Figure 2.** Typical XRD patterns at various pressures for $CH_3NH_3PbI_3$. Arrows mark the new diffraction peaks appearing at 0.3 GPa.

**Figure 3.** (a) High pressure crystal structure of $CH_3NH_3PbI_3$ viewed along the crystallographic axes c, b, and a, respectively. Hydrogen atoms are omitted for clarity. (b) Rietveld refinement of the high pressure orthorhombic phase with space group *Imm*2.

**Figure 4.** Pressure dependence of the unit cell volume of $CH_3NH_3PbI_3$ and the high pressure *Imm*2 phase is fitted by the third-order Birch–Murnaghan equation of state.

**Figure 5.** Selected IR spectra of $CH_3NH_3PbI_3$ at elevated pressures. Arrows mark the redshift and blueshift of IR peaks with increasing pressure.

**Figure 6.** *In situ* high pressure optical micrographs of $CH_3NH_3PbI_3$ in a diamond anvil cell.

**Figure 7.** Selected optical absorption and PL spectra of $CH_3NH_3PbI_3$ at different pressures. Arrows indicate the evolution of the spectra as a function of pressure.

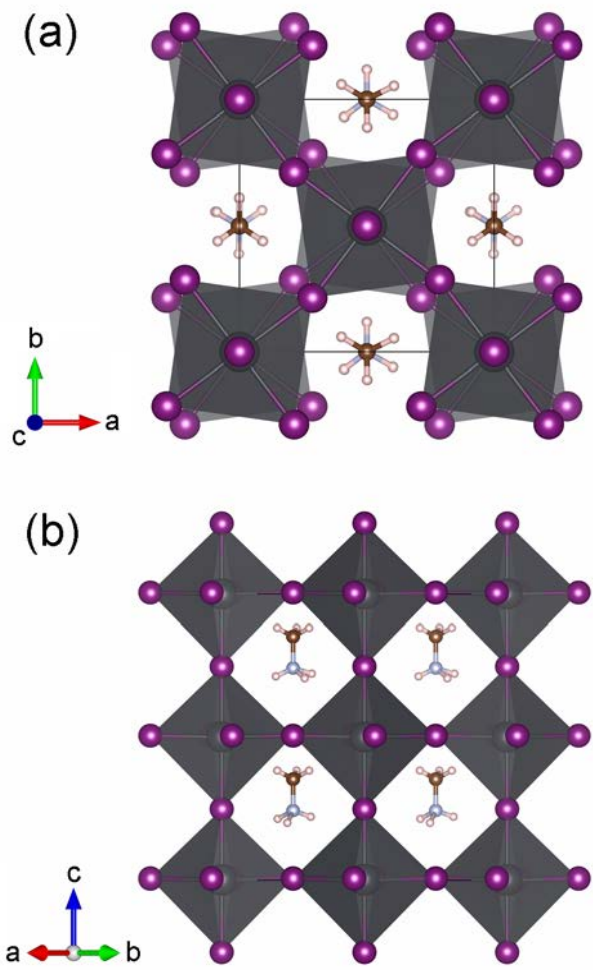

**Fig. 1**

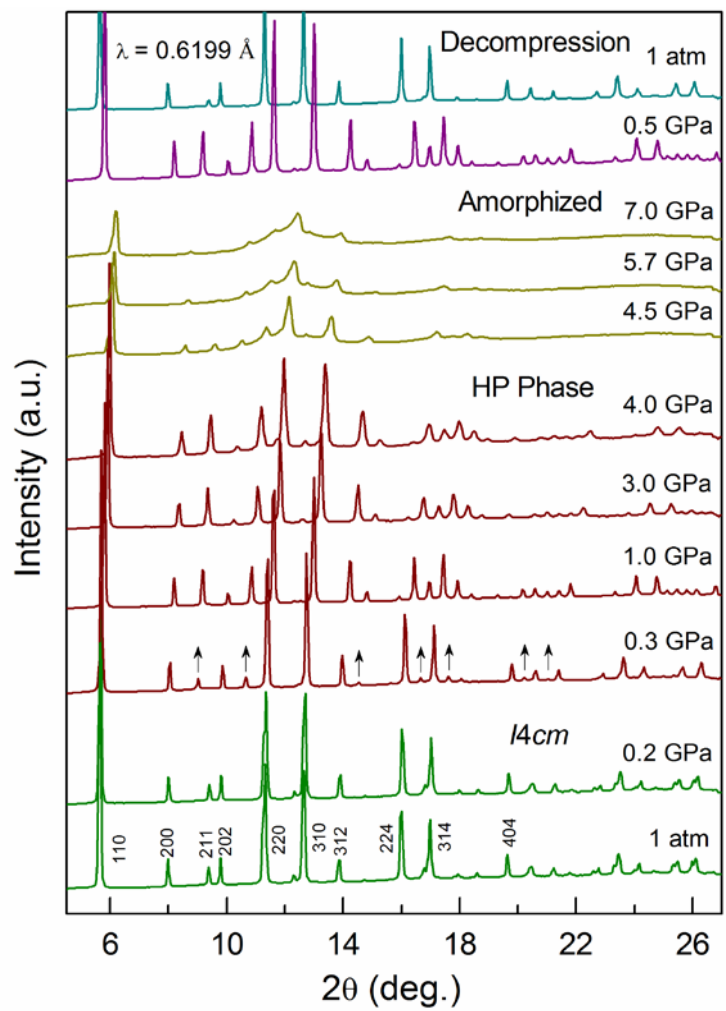

**Fig. 2**

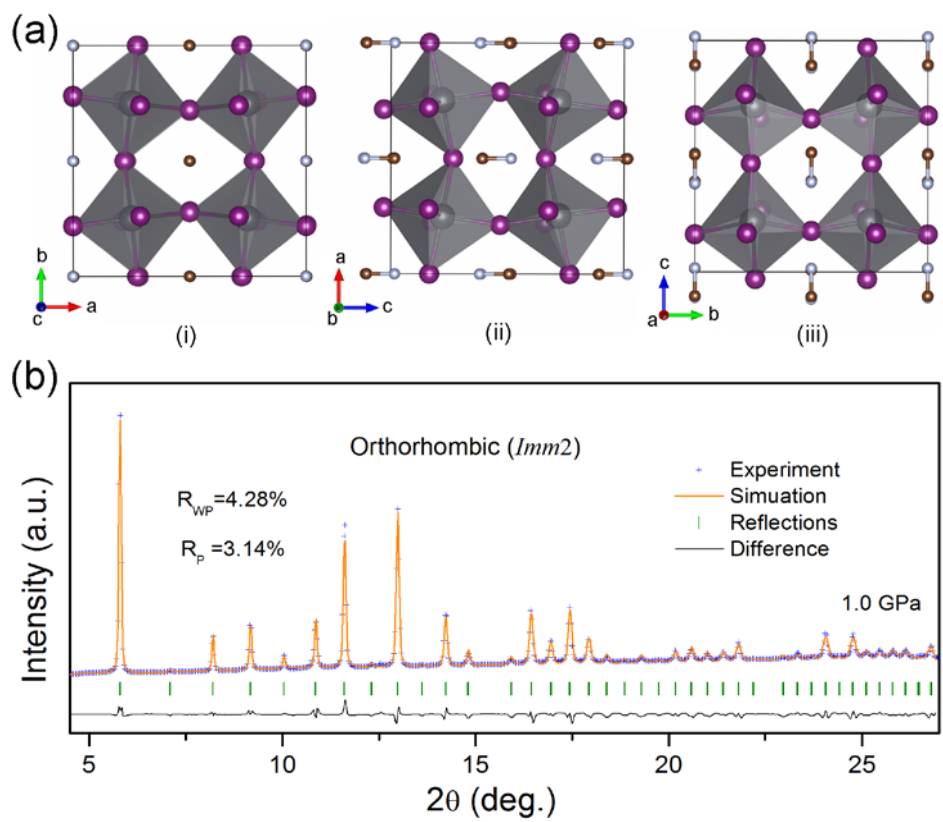

**Fig. 3**

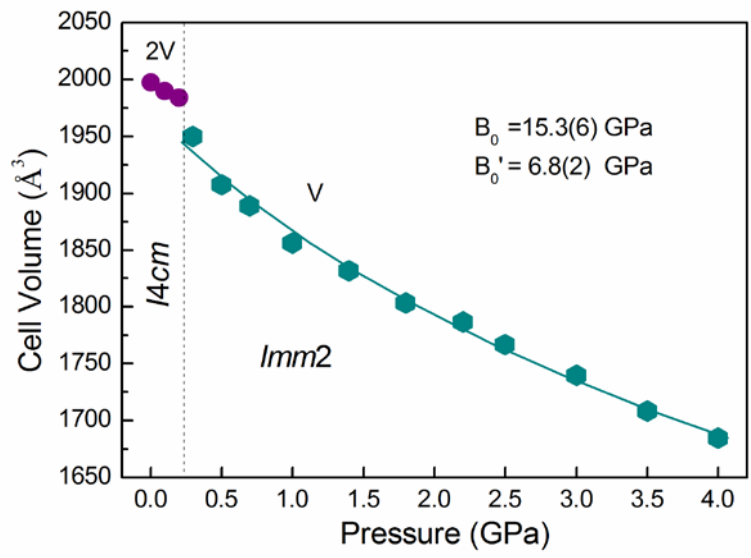

**Fig. 4**

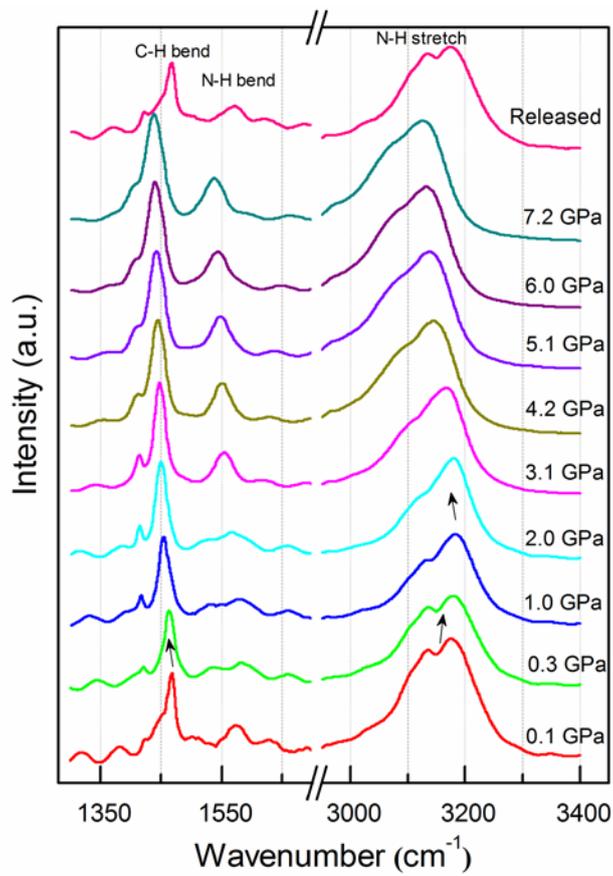

**Fig. 5**

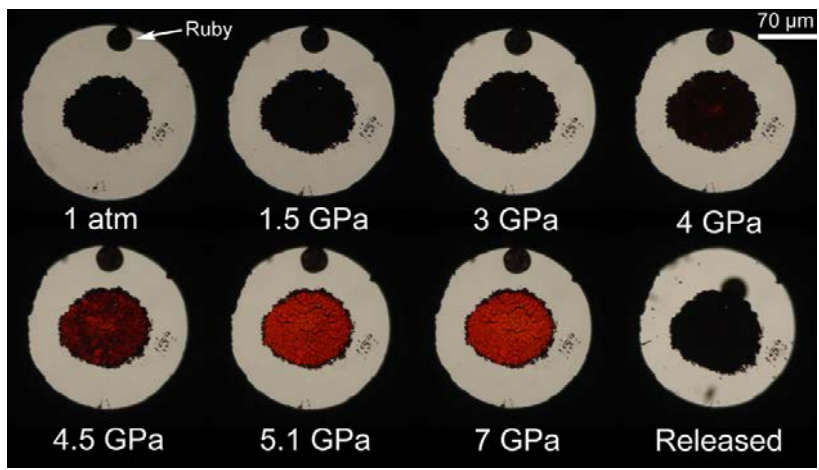

**Fig. 6**

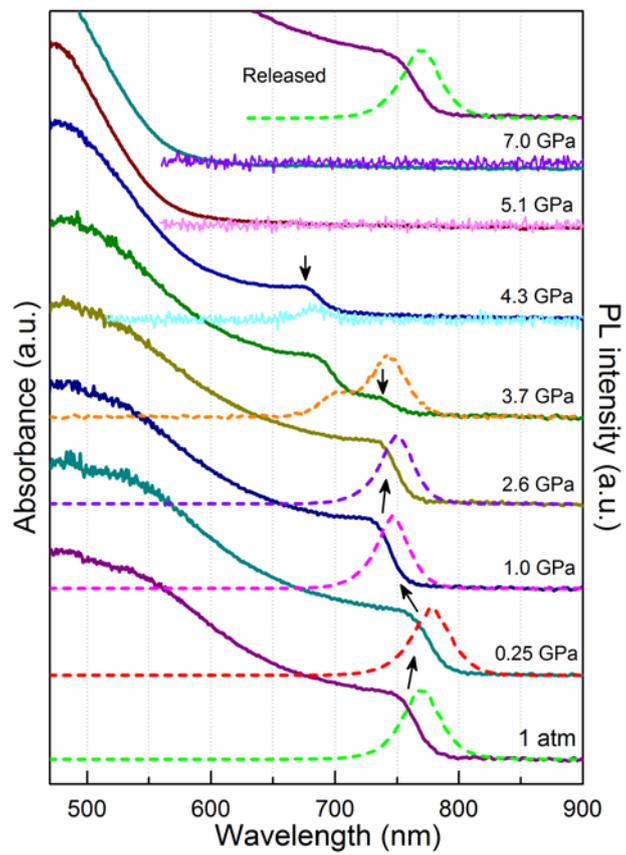

**Fig. 7**